# Exploring inorganic and nontoxic double perovskites $Cs_2AgInBr_{(6-x)}Cl_x$ from material selection to device design in material genome approach

Yunting Liang*

[1] School of Materials Science and Engineering, Zhengzhou University, Zhengzhou 450001, China

[2] State Centre for International Cooperation on Designer Low-carbon & Environmental Materials, Zhengzhou University, 100 Kexue Avenue, Zhengzhou 450001, China

[3]Zhengzhou Materials Genome Institute (ZMGI), Zhongyuanzhigu, Building 2, Xingyang 450100, China.

Corresponding author: * yunting_liang@gs.zzu.edu.cn

**Abstract:** Halide double perovskites have recently been proposed as potential environmentally friendly alternatives to organic group and lead-based hybrid halide perovskites. In particular, Cs2BiAg$X_6$ (X = Cl, Br) have been synthesized and found to exhibit tunable band gaps in the visible range. However, the band gaps of these compounds are indirect, not ideal for applications in thin film photovoltaics. Here in this work we have carried out systematic modelling, using a materials genome approach in the framework of the density functional theory (DFT), to formulate a new system of solar absorption layer based on $Cs_2InAgX_6$ and its heterojunction device. Through Cl partial substitution on Br in $Cs_2InAgBr_6$ to optimize its thermodynamic stability after the calculation of ATAT proportion searching and Gibbs free energy, we have identified a series of stable cubic-structured phases, with the general formula of $Cs_2AgInBr_{(6-x)}Cl_x$ ($0 \leq x \leq 6$), as remarkable solar absorption layer to enable harvesting the solar energy. The optimized $Cs_2InAgBr_5Cl$ compound is a marvellous solar absorption layer, with direct 1.92 eV bandgap and high solar absorption ability, consistent with $Cs_2InAgBr_6$ at standard room temperature (298 K). Assembling with anatase as n type TCO and $Cs_6Ag_4In_4Br_{18}Cl_4$ as p type TCO fabricated by introducing Cs-Br defect, the heterojunction is integrated into PSCs based on the standard p–i–n structure ($TiO_2/Cs_2InAgBr_5Cl/Cs_6Ag_4In_4Br_{18}Cl_4$), the lattice mismatching, microscopic function and band offset are evaluated for this well-designed device.

## Introduction

Exploring renewable energy is one of sustainable development strategies to solve energy crisis for human society.[1] And among these, solar energy is the most outstanding and remarkable solution without pollution to water and noise to environment, which is converted into electricity according to sunlight directly irradiating on easily installed panels on deserts and buildings. Limited by preparation cost and pollution issues in these first three generations of solar cell, organic-inorganic hybrid perovskite solar cell has achieved 23.7% conversion efficiency in less than a decade, and rivals the leading research owing to suitable bandgap, high absorption coefficient, long carrier diffusion length and reduced cost. [2-7] However, $CH_3NH_3^+$ ($MA^+$) and $HC(NH_2)_2^+$ ($FA^+$) organic groups tend to degeneration and bring about perovskite material and device instability, and Pb component causes pollution to environment according to European RoHS standard, both of which hinder its broad and commercial application for photovoltaic conversion. [8-11]

Recently, $Cs^+$ cation substituting $MA^+$ or $FA^+$ constitutes new perovskite formula $CsPbX_3$ (X=Br and I), which exhibits high-temperature endurance up to 300°C and 3 months storage cycle without sacrificing its efficiency under exposure to air. On the contrary, $MA(FA)PbI_3$ decomposes at 85°C during annealing. From this point of view, the substitution of inorganic part in $MA(FA)PbI_3$ promotes perovskite material stability, which predicts the development direction of inorganic perovskite solar cell. [12-17] Nevertheless, there is still a long way to go toward long-time stability over tens of years. Also, toxic lead is still necessary to achieve high performance. Lead pollution will do serious harm to human health, such as fatigue, muscle weakness, clumsiness and clouded consciousness. Extensive efforts have been paid to design new non-/low-toxic and stable halide perovskites for solar cells as well as other optoelectronic applications. Based on $CsPbX_3$ formula, some substitutions have been developed to replace Pb existence, e.g., bivalent ($Sn^{2+}$, $Ge^{2+}$, $Cu^{2+}$, $Sr^{2+}$), trivalent ($Bi^{3+}$, $Sb^{3+}$), tetravalent ($Sn^{4+}$) and hybrid valent (e.g. $Ag^+$ and $Bi^{3+}$).[18] But $Sn^{2+}$ ion to easily oxidize to $Sn^{4+}$, Ge raw material cost expensively, while Cu and Sr fail to realize narrow bandgap, in bivalent substitution in Pb site.[19] Trivalent and tetravalent replacements transform 3-dimensional perovskite structure (3D) with octahedral corner-sharing into 0D and 2D structures with the corresponding octahedral face-sharing and edge-sharing. Xiao et al. pointed out that 3D structure dimensionality usually caused 3D electronic dimensionality, which was related directly and positively to carrier effective mass and transport property.[20] Therefore, hybrid valent co-doping in Pb obtaining 3D double perovskite structure refers to 3+1 double cations mode, here 3 presents $Bi^{3+}$, $In^{3+}$ and $Sb^{3+}$, meanwhile 1 for $Cu^{+}$, $Ag^{+}$, $In^{+}$, $Tl^{+}$ and $Au^{+}$,[21-29] applied at multiple fields, such as solar cell, X-ray detector, photocatalysis and light-emission device et al.[30-35] Aimed at environment-friendly purpose, $Sb^{3+}$ and $Tl^{+}$ are counted as a kind of global contaminant, $Cu^{+}$ and $In^{+}$ incline to the oxidized +2 and +3 states, furthermore, $Ag^{+}$ is superior to $Au^{+}$ from material cost.[36]

Finally, $Cs_2BiAgCl_6$ have been calculated theoretically with 2.7 eV bandgap under PBE0 method, in accordance with 2.2~2.8 eV experiment report, which are beyond the gap range (1.1~1.9 eV) of a solar absorption layer.[37] After Br substituting Cl promotes valence bands upshifting, the bandgap is reduced into 2.3 eV by PBE0 calculation, compared with experimental measured 1.9~2.2 eV value. The solution-processed $Cs_2BiAgBr_6$ films exhibit excellent optoelectronic and photovoltaic properties, which enable the corresponding double perovskite solar cells to deliver a champion power conversion efficiency (PCE) of 2.51% with the highest PCE in the $Cs_2BiAgBr_6$-based solar cells to date.[38] However, the indirect bandgaps in Bi/Ag double perovskites is not ideal for thin film PV application. In fact, indirect band gaps imply weak oscillator strengths for optical absorption and for radiative recombination, therefore, indirect semiconductors such as silicon need a much thicker absorb layer, which can be problematic if charge carriers mobilities are low. Furthermore, the direct transport path in $Cs_2InAgCl_6$ reduce the transition energy consumption of photo-generated carrier, which eventually boost the PCE. However, 2.9~3.3 eV gap values by PBE0 calculation hinder the photovoltaic application of $Cs_2InAgCl_6$, in spite that its films are successfully synthesized to obtain a stable double perovskite structure with space group $Fm\bar{3}m$. Although, $Br_{Cl}$ solution could improve this wide bandgap situation, $Cs_2InAgBr_6$ has been reported to be synthesized difficultly, due to bromine accelerating the compound decomposition.[39] From this view of point, there are lack of remarkable solutions in double perovskite for PV application, which motive us to further develop more ingredient exploration.

The Materials Genome Initiative,[40] carried out in 2011 in the United States, is a largescale cooperation between materials scientists from experimentalists and theorists and computer scientists to allocate proven computational methodologies to predict, select, and optimize materials at an unprecedented scale and rate. The time for this collabration is right for this ambitious approach: it is now well comprehended that many outstanding materials properties can be predicted through solving equations based on these fundamental physics laws [41] by adopting quantum chemical approximations, for example, density functional theory (DFT). This virtual testing to select materials can be used to design and optimize materials in silico.[41,42.]

Many research units have already adopted this high-throughput computational approach to screen up innumerable compounds with potential and new technological materials. Some examples include solar photocatalysis, solar photovoltaic devices, topological insulators, photoluminescence, $CO_2$ capture materials, piezoelectric materials, and thermoelectric materials, with each research exploring several new promising candidate compounds for experimental follow-up.[43-52] Furthermore, in the fields of catalysis, hydrogen storage materials, and Li-ion batteries, experimental “hits” through high-throughput computations have already been reported.[53-57]

This work is dedicated to firstly carry out systematic theoretical simulation of high-throughput screening to explore more candidates of double perovskite, and then investigate the corresponding stability and PV properties. Furthermore, the architecture of $Cs_2InAgBr_5Cl$ based solar cells are assembled for the intention of achieving high energy conversion efficiency. First principle calculation is carried out to assess the electronic structure and band offset at the heterojunction, so as to provide reliable insight essential for improved accuracy of numerical simulation.

## Methods

Theoretical modeling calculation were carried out by using the Vienna Ab Initio Simulation Package (VASP),[58,59] while the projector augmented wave (PAW) method were adopted to describe the ionic potentials including the effect of core electrons.[60,61] In this work, the Perdew–Burke–Ernzerhof (PBE) exchange–correlation (XC) functionals[62,63] were employed to study the structural stabilities. For geometric structures relaxation, the Brillouin Zone (BZ) summation was accounted with Monkhorst–Pack k-point intervals limited below 0.04 $Å^{-1}$ for both conventional and primitive cells.[64] A plane-wave cutoff energy with 520 eV was utilized in all calculations. All obtained structures were geometrically fully relaxed until the force convergence on each ion was reduced below 0.01 eV $Å^{-1}$.[65-67] Through the calculations of electronic band structures, we performed the HSE06 functional to achieve more accurate values of band gaps.[68,69] We used a convergence criterion of $10^{-6}$ eV, adequate for electronic self-consistent iteration.

Based on the genetic evolution algorithm with the principle of energy minimization, the universal structure predictor (USPEX)[70,71] was performed to predict stable or metastable structures for any given composition. Aimed at each given composition, a population with 230 possible structures was randomly created with variable symmetries in the first generation. After the fully structure relaxation was achieved and the comparison of formation enthalpy, the most stable and metastable structures were screened into the next generation evolution. Afterwards, each subsequent generation was basically produced through heredity, with lattice mutation and permutation operators, through applying and assessing energetically for the selection of a population of 60 for next run. This calculation would continue to screen the structures from the given composition, until the most stable configuration remained unchanged for a further 20 generations to safeguard the global equilibrium.

The energy (enthalpy) of formation for each double perovskite (i.g. $Cs_2InAgBr_6$) was defined through respect to its chemical potentials of constituent phases as

$$E_f = E_c - aE_{\mathrm{CsAgBr2}} - bE_{\mathrm{AgBr}} - cE_{\mathrm{Cs3In2Br9}} \quad (1)$$

where $E_c$ is the total energy for this compound of double perovskite, and $E_{\mathrm{CsAgBr2}}$, $E_{\mathrm{AgBr}}$, and $E_{\mathrm{Cs3In2Br9}}$ are total energy for $CsAgBr_2$, AgBr, and $Cs_3In_2Br_9$, correspondingly (the ground state stable structure is used as a reference for each element/constituent as summarized in the Materials Project webpage https://www.materialsproject.org. The energy of formation with respect to that from the stable constituent phases is referred as the energy above the hull, conventionally).

The phonon frequency spectrum from a USPEX predicted structure was performed to check its dynamic stability. The supercell method concluded from the PHONOPY package[72,73] was adopted to employ the relevant frozen-phonon calculations based on harmonic approximation. The supercells expanded from relaxed structures were utilized for phonon calculations. The stability criterion is that the amplitude of imaginary frequency is less than -0.3 THz,[74,75] to tolerate degree of acceptable errors in phonon calculations.

The Alloy Theoretic Automated Toolkit (ATAT)[76,77] developed from the method of cluster expansion, is a powerful tool to build up phase equilibria of composition from the minimization of the formation energies through variable composition operation, under the frame of first-principles calculations. A cluster expansion is counted as convergence to stop, when all the conditions below are satisfied: (a) all ground states could be correctly reproduced, (b) no other new ground states are presented, and (c) the cross-validation score is typically less than 0.025 eV. Afterwards, predicted energies from the cluster expansion as a function of composition for each structure are generated. On profit from the ground states for each composition, a convex or concave hull is produced to describe the phase equilibrium.

VASPKIT was performed for the optical absorption capability for solar absorber. The frequency-dependent dielectric function $\varepsilon(\omega) = \varepsilon_1(\omega) + i\varepsilon_2(\omega)$ was calculated by using the PAW methodology, and then the absorption coefficients as a function of photon energy were obtained. Through PBE functional, the optical absorption spectra were calculated, since dense k-points (grid spacing less than $2\pi \times 0.01$ Å$^{-1}$) were required to gain a fine optical spectrum, caused high computational cost. The optical spectrum was further shifted to the HSE06-calculated bandgap to remain the spectral shape in the process. The standard AM1.5G solar spectrum at 25 °C was used to match the solar absorption range.

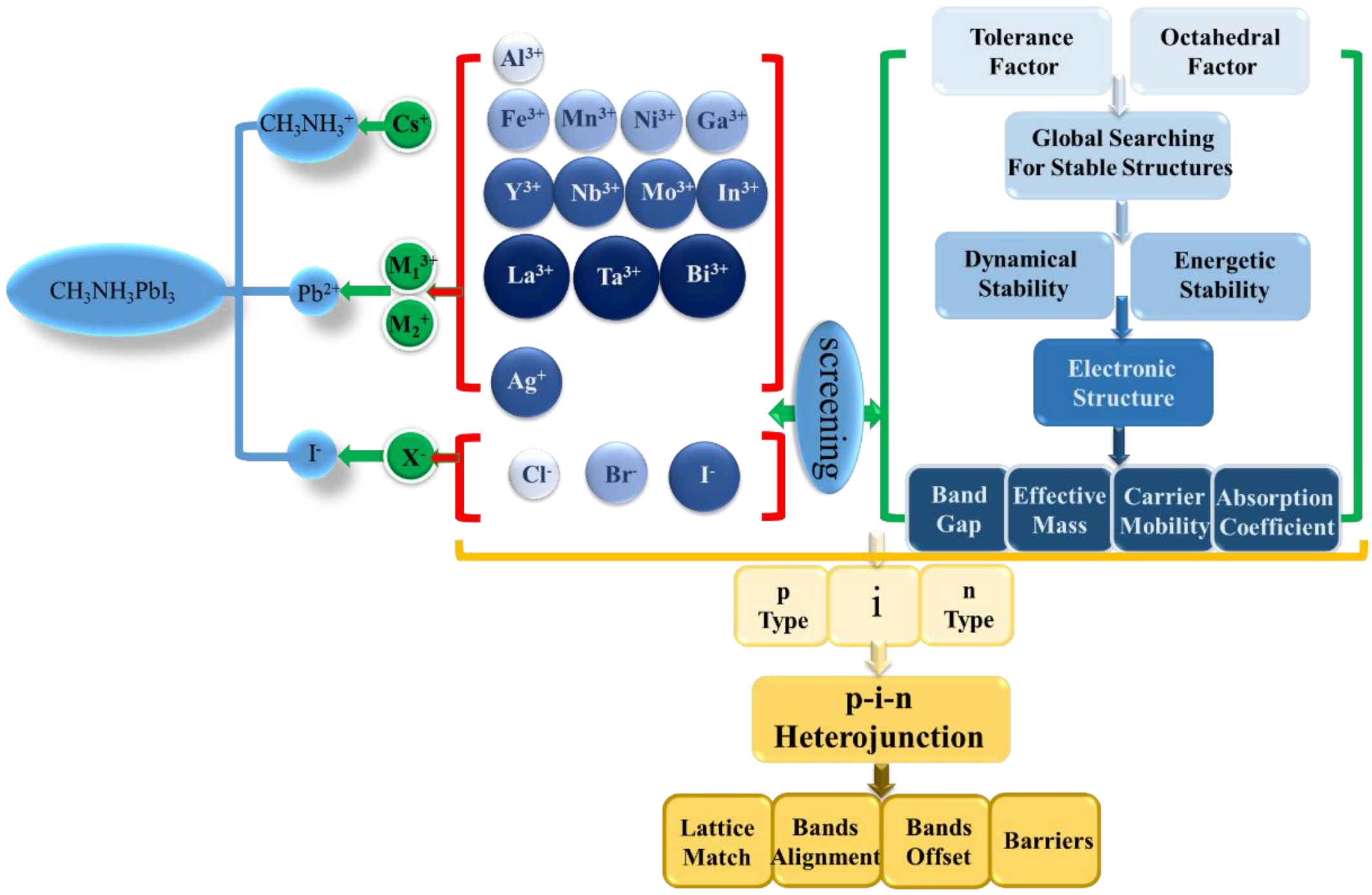

Fig.1 Flow chart for the materials genome approach.

A collection of methods treated as an integrated material genome approach for this current work was summarized and presented in Fig. 1. For each given composition, a potential stable phase can be identified as energetically and dynamically stable, when its structure is created together with a negative energy below the hull to account as the global equilibration. An allowance of 25 meV per atom can be regard as the formation energy at 0 K to tolerate the thermal fluctuation at the standard room temperature (298 K).

The variable compositional dependence on phase stability, which is typically involved in interfaces between any two terminal compositions, can be readily simulated through the ATAT code, which results in a convex or concave hull for the formation energies, therefore causing the identification on thermodynamically stable phases between the stable phases of the terminal substances. The utilization of PHONOPY simulation on thermodynamically stable structures leads to the assessment to the dynamical stability and phonon entropy as a major contribution to Gibbs free energy.

## Results and Discussion

### High-throughput screening of $Cs_2B^{3+}AgCl_6$

In order to extend potential trivalent dopant for more substituted choice, we firstly and tentatively select 12 kinds of trivalent cations in periodic table, based on two principles of resource abundant and nontoxicity or hypotoxicity, as shown in Fig. 2. These compounds are screen layer by layer, according to five types of verification method in Table S1. Firstly, 12 kinds of double perovskites are evaluated from tolerance factor and octahedron factor, for initial structure stability from Pauling rules.[78] Secondly, we adopt USPEX tool to predict the most stable structure of compounds according to the lowest enthalpy in 230 space groups, $Cs_2GaAgCl_6$ and $Cs_2AlAgCl_6$ fail to keep octahedral packing, others still remain octahedrons with Cs filling in the interspace, and the lattice parameters of stable structures are listed in Table S2. Thirdly, phonon dispersion and formation energy are utilized for stability evaluation from lattice dynamics and thermodynamics deduced from decomposition equation, separately. When large imaginary frequencies emerge below the criterion of -0.3THz for associated compound of $Cs_2MnAgCl_6$ and $Cs_2FeAgCl_6$ as shown in Fig. S2, it implies they are dynamically unstable and less likely to occur in nature, which means such structures are likely to collapse due to mechanical instability. The standard of 0.025 eV upper limit caused by thermal fluctuation distinguish $Cs_2NbAgCl_6$ and $Cs_2MoAgCl_6$ in unstable energy condition with the tendency of decomposition, which is explained from their higher stable oxidation states rather than +3 valence. Finally, only 5 types of compounds are left for the electronic structure calculation to acquire their electronic properties. The band structures of $Cs_2YAgCl_6$ and $Cs_2LaAgCl_6$ exhibit semiconductor character with >3.0 eV wide bandgap, failing to become transparent conductive oxides (TCOs) with the distance between VBM and fermi level larger than 0.026 eV thermal excitation limit. The same situation happens in $Cs_2NiAgCl_6$ with impurity bands existing in forbidden gap. Only $Cs_2BiAgCl_6$ and $Cs_2InAgCl_6$ fulfill the photovoltaic requirement as potential choices for inorganic double perovskite solar cell, which are in accordance with previous works. They are worth to further investigate from our validated workflow in line with the idea from material genome project.

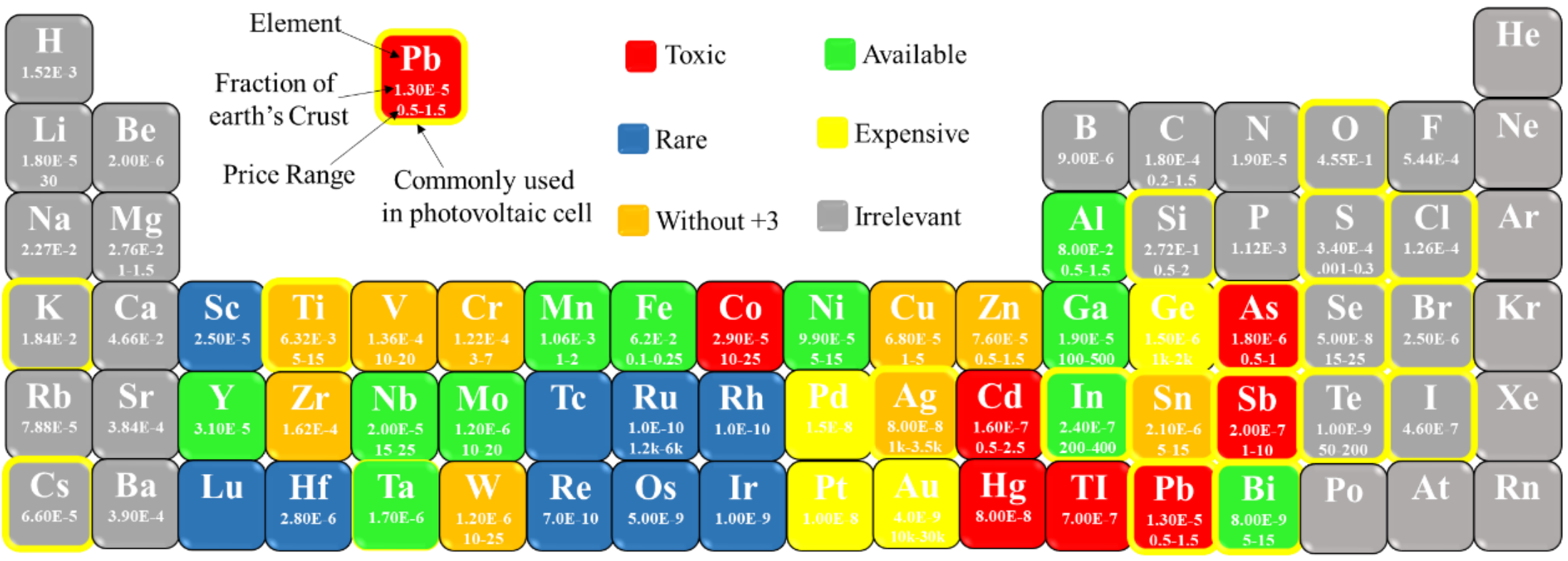

Fig.2 trivalent cation selection in periodic table according to resource abundant and nontoxicity or hypotoxicity.

More accurate band structures of $Cs_2InAgCl_6$ and $Cs_2BiAgCl_6$ are obtained under HSE06 methods, because of GGA usually underestimating their gap values lack of considering the non-local effect, as shown in Fig. 3 (a) and (b). Although the indirect bandgap of $Cs_2BiAgCl_6$ along X to L point is 0.78eV lower than the direct gap of $Cs_2InAgCl_6$ at G-G point, the indirect bandgap infers that carrier transport along indirect intra-band of $Cs_2BiAgCl_6$ generate photoluminescence effect, which consume carrier energy to cause carrier annihilation and reduce the PV conversion efficiency. Inversely, the direct bandgap in $Cs_2InAgCl_6$ achieve PV quantum effect in inter-bands for higher efficiency. However, after HSE06 correction, 3.3 eV bandgap in $Cs_2InAgCl_6$ are largely beyond the absorption range to visible light. Fortunately, the improvement on the gap reduction has been realized in $Cs_2InAgBr_6$ from previous works,[39] by the bigger radius Br substituting Cl atoms to upshift valence bands, and its band structure with 1.64 eV bandgap as shown in Fig. 3 (c).

Figs. 3(d)~(e) are calculated phonon band structures for $Cs_2InAgCl_6$ and $Cs_2BiAgCl_6$ $Cs_2InAgBr_6$, where the grey dotted lines are treated as zero frequency. The cyan lines are indicated as the acceptable stability margin of -0.3 THz in phonon calculations, which comes from the high sensitivity of the fully-relaxed structures to remnant stress. Through checking the characters of their phonon band structures, the dynamic stabilities for these relaxed structures from USPEX searched energetically stable configuration at 0 K are evaluated. When no phonon bands are connected with imaginary frequencies, it predicts that this phase is dynamically stable. It is seen from Fig. 3 that the margin of these three relaxed structures are far from -0.3 THz, which demonstrates that all these structures can be considered to be dynamically stable.

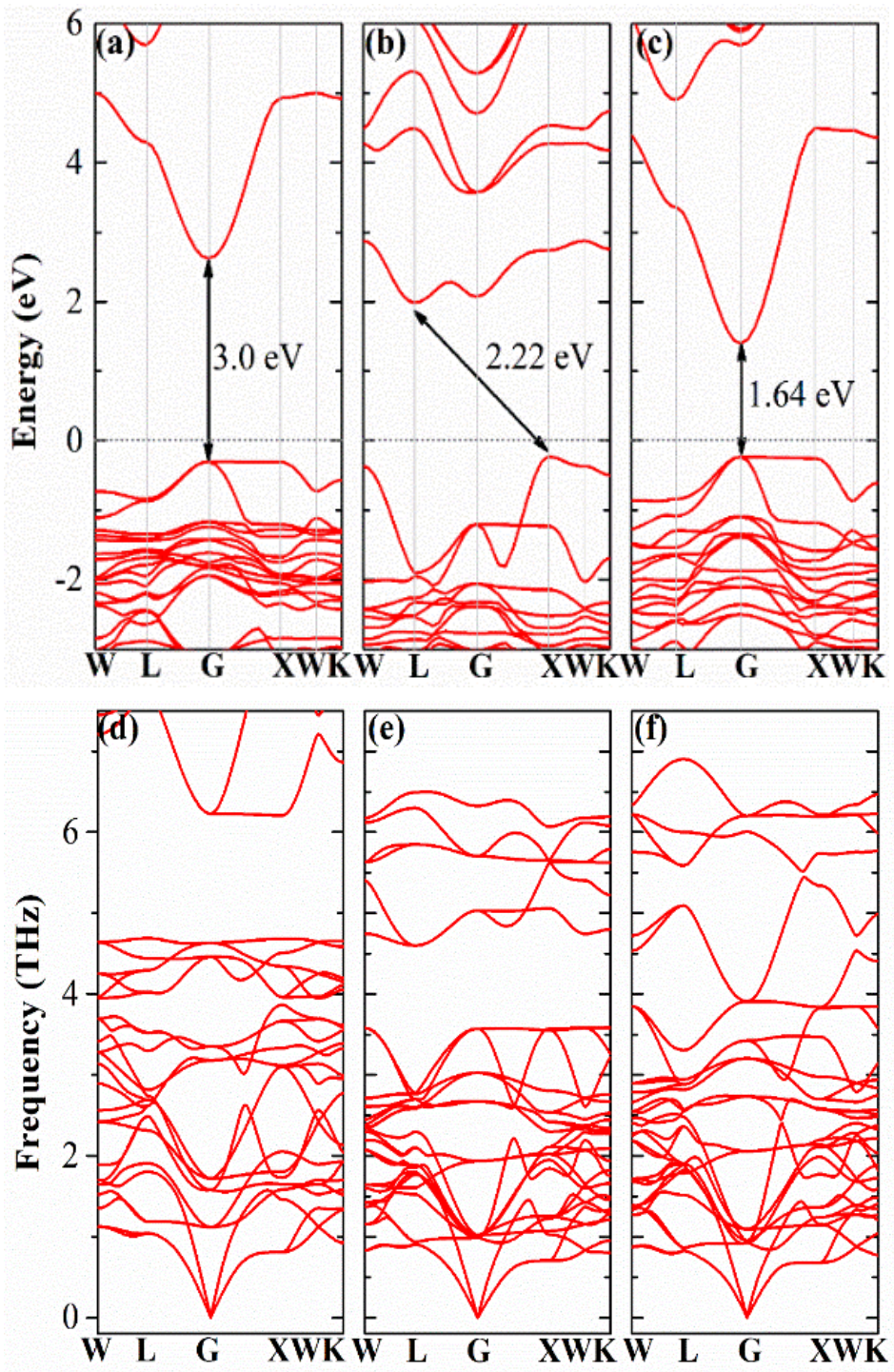

Fig. 3 (a) The band structures of $Cs_2InAgCl_6$ and $Cs_2BiAgCl_6$ $Cs_2InAgBr_6$; (b) The corresponding phonon band structures.

## The thermodynamics stability of formation energy

We then examine their thermodynamics stability of three compounds ($Cs_2AgInCl_6$, $Cs_2AgBiBr_6$ and $Cs_2AgInBr_6$) to identify their ternary compounds with their corresponding stable constituent phases.[79] We firstly build up their pseudo-ternary phase diagrams at 0K, as shown in Fig.4 (a)~(c), according to their potential constituent candidates selected in the Materials Project database, and their corresponding formation energies. The dotted lines showed in pseudo-ternary phase diagrams of $Cs_2AgBiBr_6$ and $Cs_2AgInBr_6$ at 0K predict that they are less stable with respect to their constituent phases of $Cs_2AgBr_3$, $Cs_3B_2Br_9$ (B=Bi, In) and AgBr, with $Cs_2AgInCl_6$ being more stable than $Cs_2AgBiBr_6$ and $Cs_2AgInBr_6$. Because of their formation energies being less negative than the sum results from linear combinations of their respective constituent phase, these two compounds energetically incline to decompose into these constituent components below at the ground state:

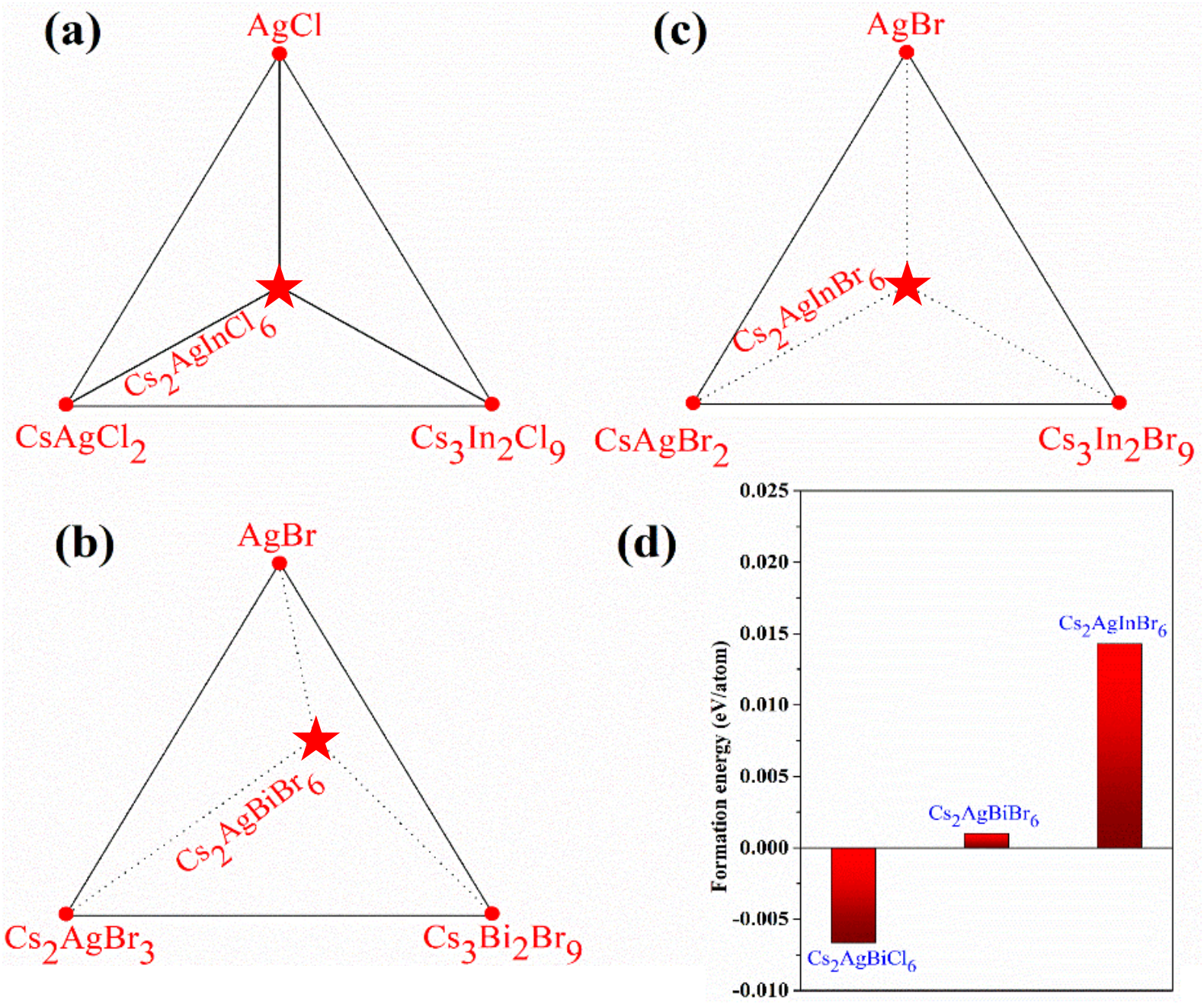

Fig. 4 (a)~(c) Ternary diagrams and formation energies for $Cs_2AgInCl_6$, $Cs_2AgBiBr_6$ and $Cs_2AgInBr_6$ based alloys with respect to stable constituent phases; (d) the corresponding formation energy.

$$Cs_2AgBiBr_6=0.25Cs_2AgBr_3+0.75AgBr+0.5Cs_3Bi_2Br_9 \quad (2)$$

$$Cs_2AgInBr_6=0.5CsAgBr_2+0.5AgBr+0.5Cs_3In_2Br_9 \quad (3)$$

In Fig. 4 (d), the corresponding formation energies for these three compounds are respectively -0.0066, 0.000986 and 0.0142 eV per atom. Normally, the allowance of 25 meV per atom can be considered as the upper limit for the formation energy at 0 K to accommodate the thermal fluctuation at standard room temperature (298 K). Besides, it is reported that $Cs_2BiAgBr_6$ films have been prepared experimentally with 2.51% PCE,[38] and $Cs_2InAgBr_6$ fail to be synthesized.

**ATAT proportion searching for $Cs_2AgInBr_{(6-x)}Cl_x$**

Although $Cs_2AgInBr_6$ tunes the bandgap value under 1.9 eV for PV application, its formation energy reveals it is unable to obtain from experimental synthesis. Previous work suggests that the stability improvement of Br substitution in $MAPbBr_xI_{1-x}$ solution provide a new perspective to increase material stability.[80] Under the inspiration of this idea, we perform $Cs_2AgInBr_{(6-x)}Cl_x$ with Cl partial substitution to modify the parent stability, at the same time, and keep the direct and tunable bandgap.

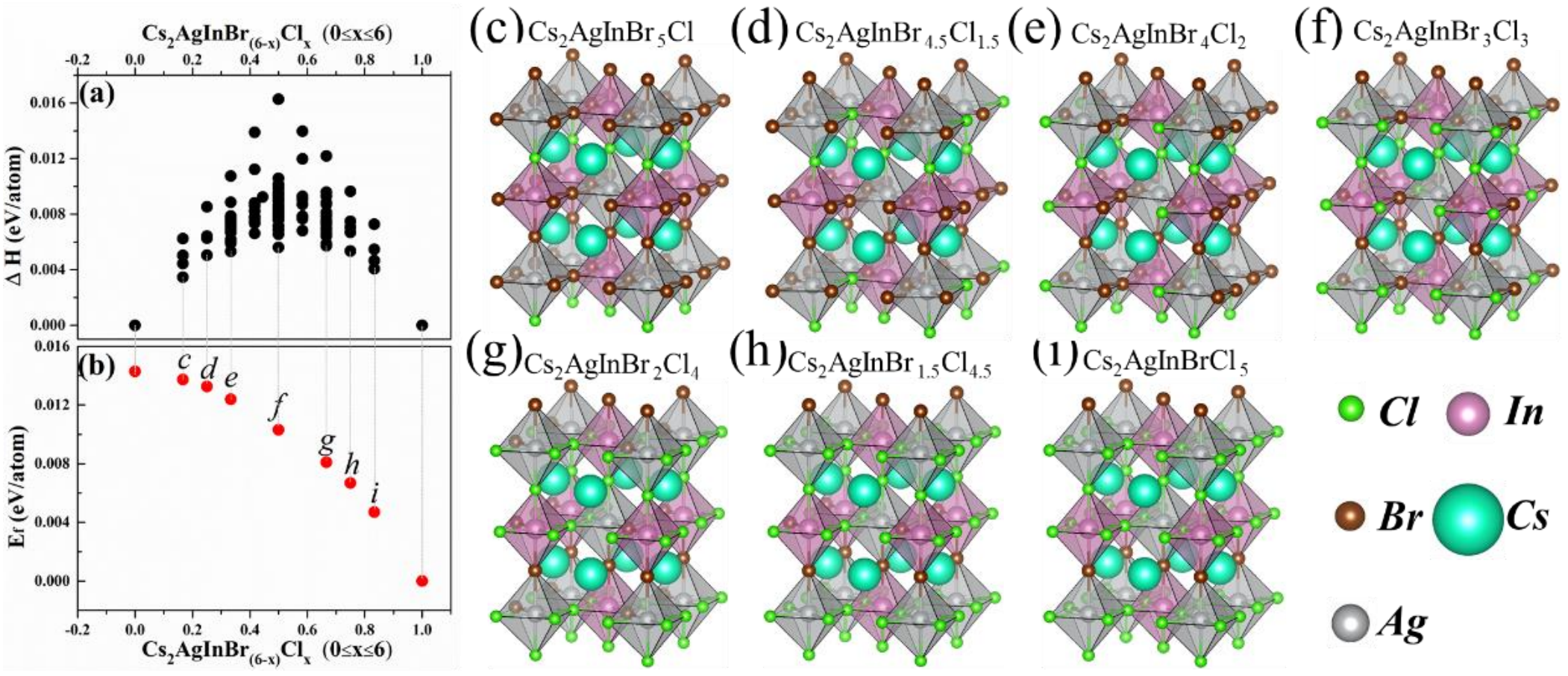

Fig. 5 (a) The convex hull for $Cs_2AgInBr_{(6-x)}Cl_x$ ($0 \le x \le 6$) through ATAT simulation. (b) Stable structure for $Cs_2AgInBr_{(6-x)}Cl_x$ corresponding to minimal energy.

ATAT searching outcome for the stable configurations of Cl-Br ratio is represented as a concave hull, through the combination of DFT with cluster expansion simulation over $Cs_2AgInBr_{(6-x)}Cl_x$. As is shown in Fig. 5 (a), these most stable structural configurations are based on Cl substitution of variable ratio (x=1, 1.5, 2, 3, 4, 4,5 and 5). On the profit from the principle of chemical potential minimization, their corresponding stable phases are those sitting at the bottom edge of the concave hull over the minimal common tangents between two limit points. And the stable phases at the bottom of the hull are further investigated according to their thermodynamic stability. With the increasing Br content, the formation energy of $Cs_2AgInBr_{(6-x)}Cl_x$ arises, but still below the thermal fluctuation limit of 25 meV per atom, which means they can survive during heating, as shown in Fig .5 (b). Fig. 5 (c)~(i) present their corresponding crystal structures of different Cl-Br ratio, with $AgX_6$ and $InX_6$ (X=Br, Cl) octahedrons alternately connecting and Cs atoms filling in the interspace, and all structures keep cubic phase with PM3M symmetry.

**Gibbs free energy for the stability of variable temperature of $Cs_2InAgBr_{(6-x)}Cl_x$**

We further analysize the thermal influences on Gibbs free energies of these compounds within the quasi-harmonic approximation to consider the phonon entropy, shown in Fig. 6 (a). Through its Gibbs calculation, $Cs_2InAgBr_6$ is indeed demonstrated as energetically unstable over a wide temperature range. However, according to other calculated Gibbs free energies in the range from 150 to 184K, seven kinds of $Cs_2InAgBr_{(6-x)}Cl_x$ is stable with respect to their decomposition products, with their formation energy getting more negative against the increasing temperature. As temperature rises to 150K, $Cs_2InAgBrCl_5$ turns to be energetically stable with respect to $CsAgCl_2$, $Cs_3In_2Cl_9$ and AgBr, while $Cs_2InAgBr_5Cl$ is still stable with reference to $CsAgBr_2$, $Cs_3In_2Br_9$ and AgCl over 184K. In spite that positive formation energy in Fig. 4 (d) predict their thermodynamic stability at 0K, $Cs_2InAgBr_{(6-x)}Cl_x$ ($0 \leq x \leq 6$) can exist during the temperature rising. The corresponding band structures of different Cl-Br ratio are listed in Fig. 6 (b). With the increasing Cl content, the CBM upshifts gradually, while the VBM keeps almost the same position. Cl substitution extend the bandgap by 0.71 eV from x=1 to x=5, without changing the band morphology.

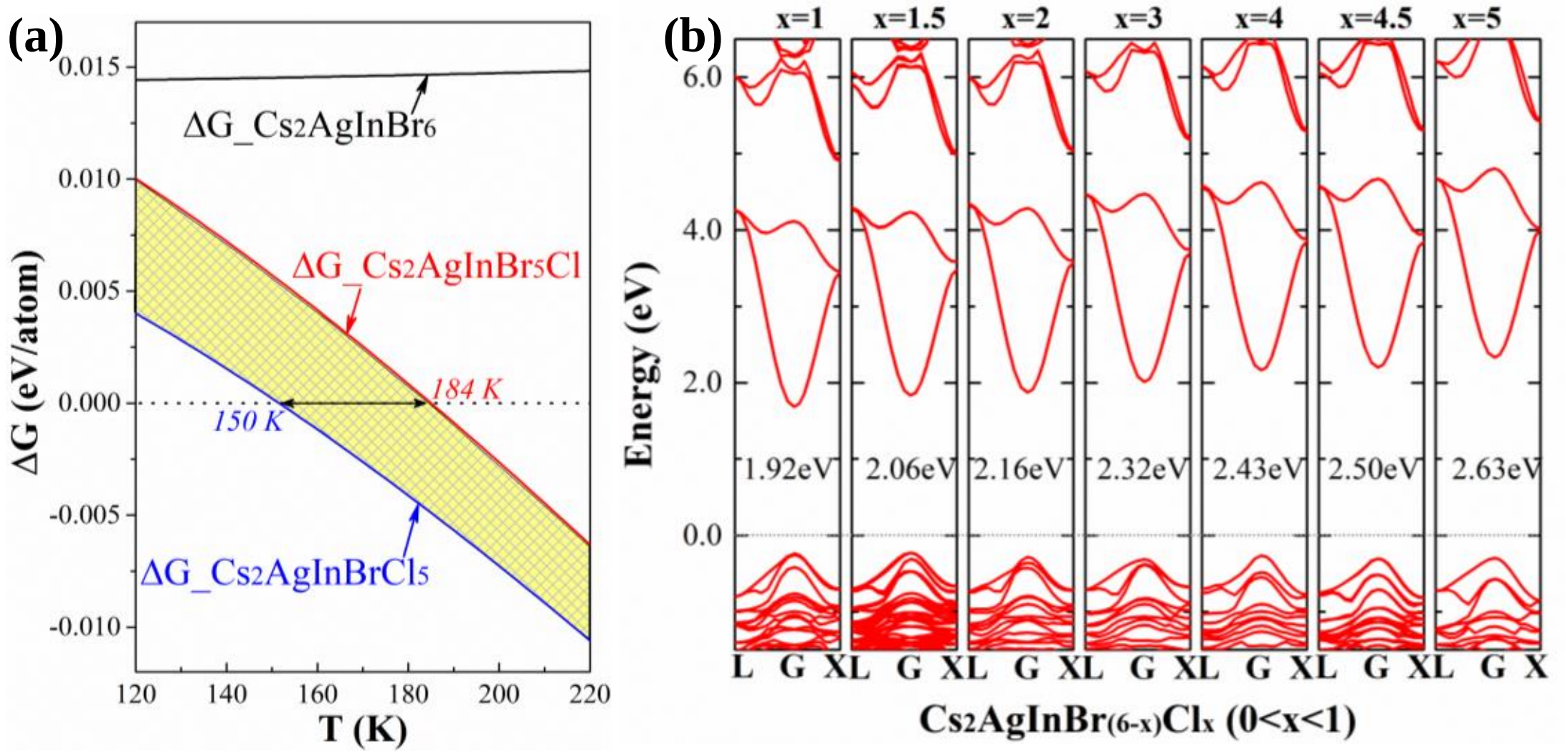

Fig. 6 (a) The stability of variable temperature of $Cs_2AgInBr_{(6-x)}Cl_x$ from AIMD simulation; (b) The corresponding band structures of $Cs_2AgInBr_{(6-x)}Cl_x$.

## Optical absorption coefficient of $Cs_2InAgBr_5Cl$

The absorption coefficient of $Cs_2AgInCl_6$, $Cs_2AgInBr_6$ and $Cs_2AgInBr_5Cl$ are calculated to evaluate their absorption ability on visible light with the range from

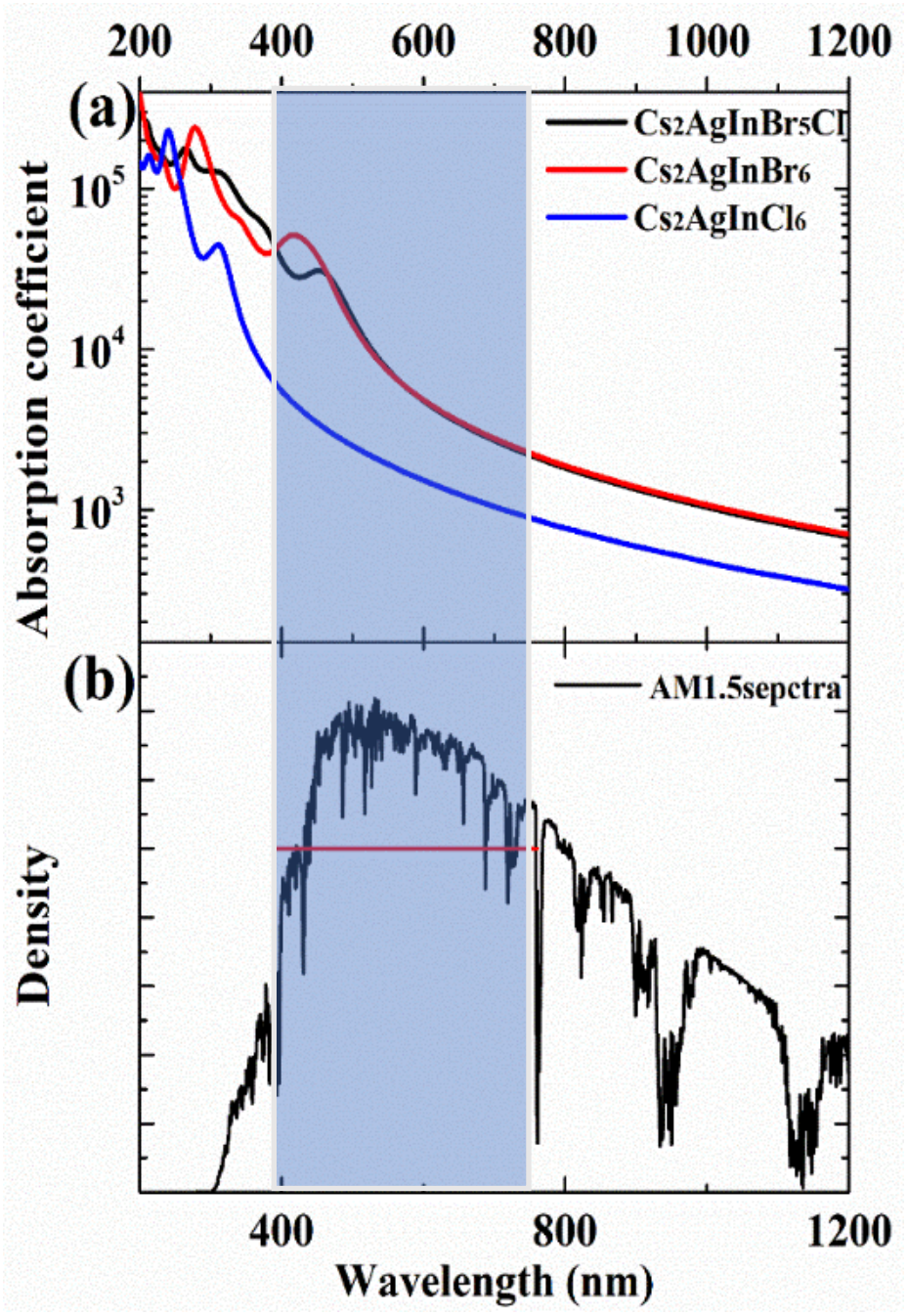

Fig. 7 (a) Optical absorption ability of $Cs_2AgInCl_6$, $Cs_2InAgBr_6$ and $Cs_2AgInBr_5Cl$; (b) The standard AM1.5G solar spectrum at 25 °C.

390nm to 760nm, as the blue rectangle shown in Fig. 7 (b). We find 3.3 eV wide bandgap in $Cs_2AgInCl_6$ causes worse absorption coefficient, meanwhile, $Cs_2AgInBr_6$ and $Cs_2AgInBr_5Cl$ achieve almost the same coefficient, which means Cl substitution in $Cs_2AgInBr_5Cl$ obtaining the same optical property with $Cs_2AgInBr_6$, displayed in Fig. 7 (a).

## The heterojunction construction

The above work concludes that $Cs_2AgInBr_5Cl$ is a remarkable solar absorption layer with the thermodynamic stability and 1.92 eV direct bandgap. The subsequent work is attempted to construct the generated PV heterojunction with $Cs_2AgInBr_5Cl$ as perovskite absorber. On contrast with rutile with 3.0 eV bandgap, 3.2 eV anatase ($TiO_2$) enable the visible light maximumly penetrating into light absorber and then acquire higher open-circuit voltage, playing as n type transparent conductive oxide (TCO), and its crystal structure is displayed in Fig. S2 (a). Furthermore, based on ATAT proportion and structure searching, introducing Cs-Br defect in $Cs_{2-x}(Agln)Br_{5-x}Cl$ compound produces $Cs_6Ag_4In_4Br_{18}Cl_4$ with p type TCO character, and the crystal structure and band structure of $Cs_6Ag_4In_4Br_{18}Cl_4$ are exhibited in Fig. S2 (b) and (c), respectively.

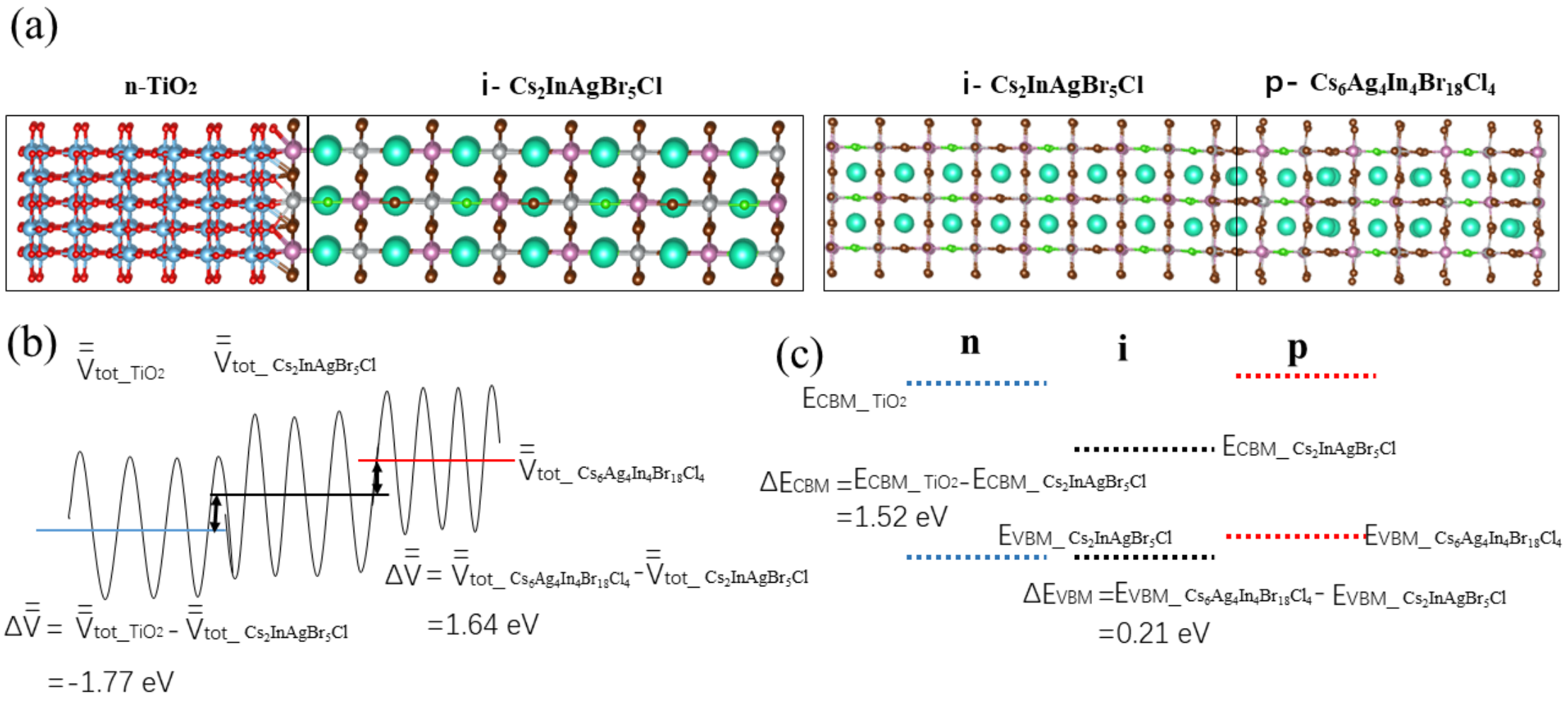

Fig. 8. (a) The atomic configurations and lattice matching on the interface of heterostructure composed of $TiO_2/Cs_2InAgBr_5Cl$ and $Cs_2InAgBr_5Cl/Cs_6Ag_4In_4Br_{18}Cl_4$ slabs; schematic summary of microscopic difference in n-i and i-p heterostructure (b) and band offset (c) from DFT calculation.

The heterojunction model with n–i–p structure ($TiO_2/Cs_2InAgBr_5Cl/Cs_6Ag_4In_4Br_{18}Cl_4$) is built through full structural relaxation in DFT calculation using the GGA functional. The resultant stable structure is of fundamentally coherent lattice matching between $TiO_2$ (1 0 1) and $Cs_2InAgBr_5Cl$ (0 0 1) planes, and between $Cs_2InAgBr_5Cl$ (0 0 1) and $Cs_6Ag_4In_4Br_{18}Cl_4$ (1 0 0), with both of lattice mismatching less than 6%. Fig. 8(a) demonstrates the excellent chemical bonding in both interfaces. A vacuum layer over 1 nm is used to avoid interaction between images of the coupled $TiO_2/Cs_2InAgBr_5Cl$ and $Cs_2InAgBr_5Cl/Cs_6Ag_4In_4Br_{18}Cl_4$ slabs. For the accurate calculation of energy band structures, the non-local effect in the exchange-correlation functional is developed by using the HSE hybrid functional. It concludes that w=0.2 screen parameter is adopted to produce their corresponding and reliable band gap values. In order to determine the average potentials of a hetero-structure constructed by these three bulk materials, the microscopic average potential can be counted as pseudopotential calculation by averaging the sum of the exchange-correlation, and ionic potentials parallel to the interface, defined as $\bar{\bar{V}}tot(z)$, a one-dimensional quantity along the perpendicular direction to the interface.

The differences in potential across these two interfaces of this heterostructure can be defined as:

$$\Delta\bar{\bar{V}}tot = \bar{\bar{V}}tot_{TiO_2}(z) - \bar{\bar{V}}tot_{Cs_2InAgBr_5Cl}(z) \quad (4)$$

$$\Delta\bar{\bar{V}}tot = \bar{\bar{V}}tot_{Cs_6Ag_4In_4Br_{18}Cl_4}(z) - \bar{\bar{V}}tot_{Cs_2InAgBr_5Cl}(z) \quad (5)$$

For the offset from the macroscopic average potentials, $\Delta\bar{\bar{V}}tot$ is however inadequate to describe the behavior with moving electrons across the heterojunction, as the maximum of CBM and VBM from bulk materials for each side of interface should be taken into account. After considering the valence band and conduction band, the corrected band offsets are given as below:

$$\Delta E_C = (\bar{\bar{V}}tot_{TiO_2} - \bar{\bar{V}}tot_{Cs_2InAgBr_5Cl}) + (E_{CMB TiO_2} - E_{CMB Cs_2InAgBr_5Cl}) \quad (6)$$

$$\Delta E_V = (\bar{\bar{V}}tot_{Cs_6Ag_4In_4Br_{18}Cl_4} - \bar{\bar{V}}tot_{Cs_2InAgBr_5Cl}) + (E_{VMB Cs_6Ag_4In_4Br_{18}Cl_4} - E_{VMB Cs_2InAgBr_5Cl}) \quad (7)$$

Where $\Delta E_{CBM}$ and $\Delta E_{VBM}$ are the difference in the energies of conduction band minimum from the bulk $TiO_2$, $Cs_2InAgBr_5Cl$, and the energies of valence band maximum from the bulk $Cs_2InAgBr_5Cl$ and $Cs_6Ag_4In_4Br_{18}Cl_4$. Here we follow up the convention way in assigning signs to the valence and conduction band offset with respect to $Cs_2InAgBr_5Cl$, such that downward offset is negative for $\Delta E_C$ but positive for $\Delta E_V$, owing to different signs for electrons and holes. From the calculation in Fig. 8 (b) and (c), the $\Delta E_C$ of n-i junction is -0.25 eV barrier with realizing the tendency of electron transport downwards, while the $\Delta E_V$ of i-p junction is 1.85 eV barrier, which predicts the hole rising. The data of band offset suggest this constructed heterojunction can work well for PV enrichment. The CBMs and VBMs of n-i-p structures in Fig. 8 (c) conclude the band alignment of this architecture belongs to type-II heterojunction, which is beneficial for manufacturing photoelectrical detectors and photovoltaic devices, owing to the advantage of the separation of photo-generated carriers, on contrast with type-I heterojunction in favour of the recombination of carriers. This perspective of band alignment get agreement with the results of band offset, demonstrating the potential candidate for PV device from this heterojunction.

## Conclusions

Systematic modelling has been carried out to design solar absorption layer of lead-free and inorganic double perovskite through optimizing the Br-Cl ratio in $Cs_2AgInBr_{(6-x)}Cl_x$ to improve its thermodynamic stability and retain the suitable direct bandgap for PV application. This results in the discovery of $Cs_2InAgBr_5Cl$, which exerts the stability from the evaluation of formation energy and Gibbs free energy, thus permitting 1.92 eV direct bandgap and the same solar absorption ability with $Cs_2InAgBr_6$ at standard room temperature (298 K).

The subsequent work is attempted to construct the PV generated heterojunction with $Cs_2AgInBr_5Cl$ as perovskite absorber. The well-designed heterojunction of p–i–n structure ($TiO_2/Cs_2InAgBr_5Cl/Cs_6Ag_4In_4Br_{18}Cl_4$) is evaluated by <6% lattice mismatching of both interfaces, expected band offset and band alignment in favour of the separation of photo-generated carriers.

**Notes**

The authors declare no competing financial interest.

**Acknowledgements**

This work is launched under the 1000 Talents Program of China, Zhengzhou Materials Genome Institute, the National Natural Science Foundation of China (No. 51001091, 111174256, 91233101, 51602094, and 11274100), and the Fundamental Research Program from the Ministry of Science and Technology of China (No. 2014CB931704).

## Supporting information

Table S1 Layer-by-layer screening process of 12 components (√ for passed, × for failed, -- for terminated)

| | Factors | USPEX | PHONO | ENERGY | BAND |
|---|---|---|---|---|---|
| $Cs_2AlAgCl_6$ | √ | × | -- | -- | -- |
| $Cs_2MnAgCl_6$ | √ | √ | × | -- | -- |
| $Cs_2FeAgCl_6$ | √ | √ | × | -- | -- |
| $Cs_2NiAgCl_6$ | √ | √ | √ | √ | √ |
| $Cs_2GaAgCl_6$ | √ | × | -- | -- | -- |
| $Cs_2YAgCl_6$ | √ | √ | √ | √ | √ |
| $Cs_2NbAgCl_6$ | √ | √ | √ | × | -- |
| $Cs_2MoAgCl_6$ | √ | √ | √ | × | -- |
| $Cs_2InAgCl_6$ | √ | √ | √ | √ | √ |
| $Cs_2LaAgCl_6$ | √ | √ | √ | √ | √ |
| $Cs_2TaAgCl_6$ | √ | √ | × | -- | -- |
| $Cs_2BiAgCl_6$ | √ | √ | √ | √ | √ |

Table S2 Lattice constants (Å, °) and symmetry of 12 components

| | a | b | c | α | β | γ | Symmetry | Space group | Phase |
|---|---|---|---|---|---|---|---|---|---|
| $Cs_2AlAgCl_6$ | 7.328 | 10.382 | 6.251 | 89 | 115 | 81 | 1 | P1 | triclinic |
| $Cs_2MnAgCl_6$ | 7.243 | 7.243 | 7.243 | 60 | 60 | 60 | 225 | $F$M$\bar{3}$M | cubic |
| $Cs_2FeAgCl_6$ | 7.211 | 7.211 | 7.211 | 60 | 60 | 60 | 225 | $F$M$\bar{3}$M | cubic |
| $Cs_2NiAgCl_6$ | 7.257 | 7.257 | 7.257 | 60 | 60 | 60 | 225 | $F$M$\bar{3}$M | cubic |
| $Cs_2GaAgCl_6$ | 6.582 | 7.54 | 8.178 | 96 | 99 | 82 | 1 | P1 | triclinic |
| $Cs_2YAgCl_6$ | 7.663 | 7.663 | 7.663 | 60 | 60 | 60 | 225 | $F$M$\bar{3}$M | cubic |
| $Cs_2NbAgCl_6$ | 7.463 | 7.463 | 7.463 | 60 | 60 | 60 | 225 | $F$M$\bar{3}$M | cubic |
| $Cs_2MoAgCl_6$ | 7.392 | 7.392 | 7.392 | 60 | 60 | 60 | 225 | $F$M$\bar{3}$M | cubic |
| $Cs_2InAgCl_6$ | 7.538 | 7.538 | 7.538 | 60 | 60 | 60 | 225 | $F$M$\bar{3}$M | cubic |
| $Cs_2LaAgCl_6$ | 7.848 | 7.848 | 7.848 | 60 | 60 | 60 | 225 | $F$M$\bar{3}$M | cubic |
| $Cs_2TaAgCl_6$ | 7.843 | 7.843 | 4.91 | 90 | 90 | 90 | 123 | P4/MMM | tetragonal |
| $Cs_2BiAgCl_6$ | 7.736 | 7.736 | 7.736 | 60 | 60 | 60 | 225 | $F$M$\bar{3}$M | cubic |

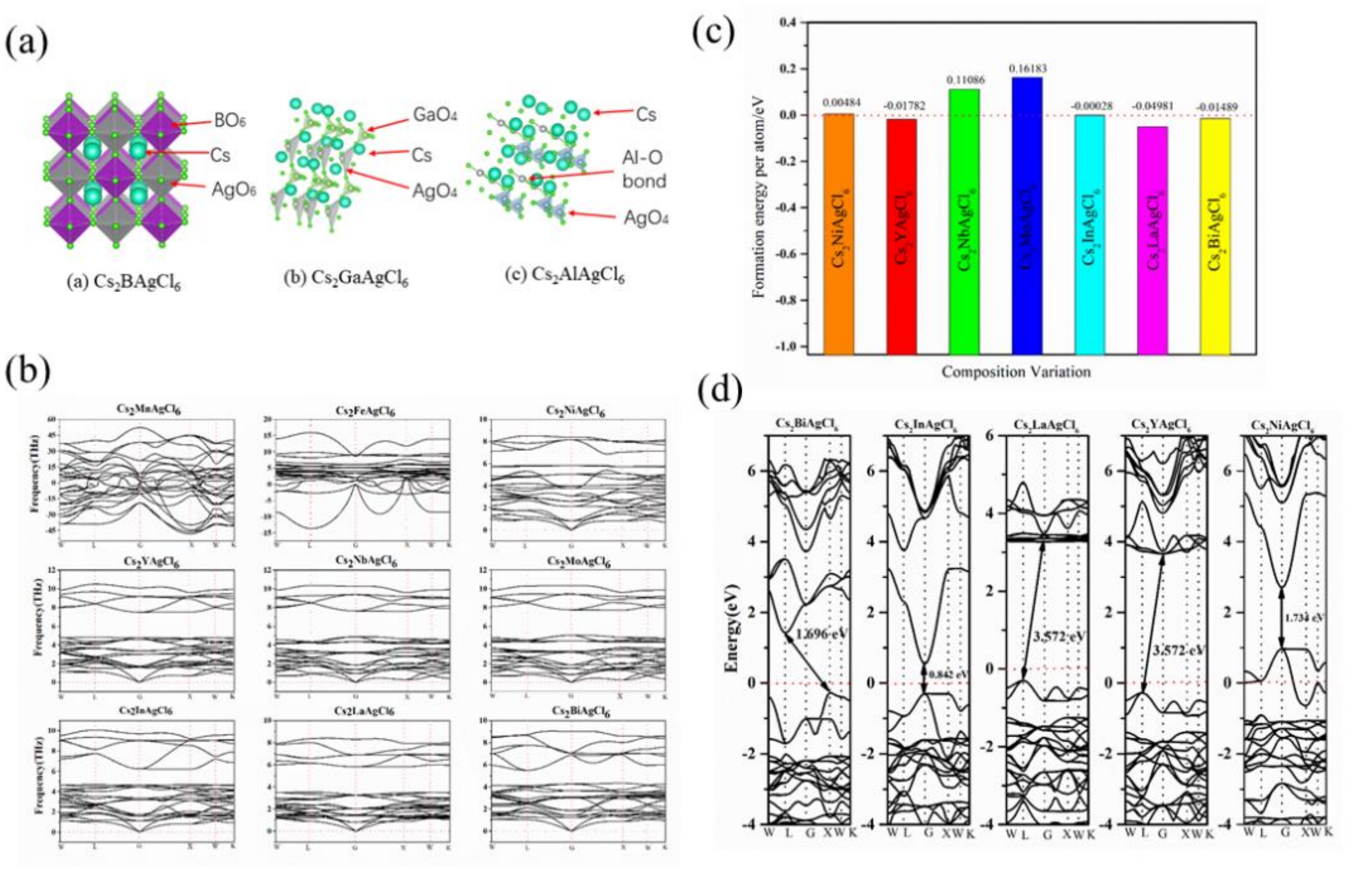

Fig.S1. (a) Crystal structures for regular cubic perovskites, $Cs_2GaAgCl_6$ and $Cs_2AlAgCl_6$; (b) phonon dispersion of 9 kinds of left components after USPEX structure screening; (c) the comparison of the formation energy of 7 kinds of left components; (d) PBE band structures of 5 kinds of preferred components.

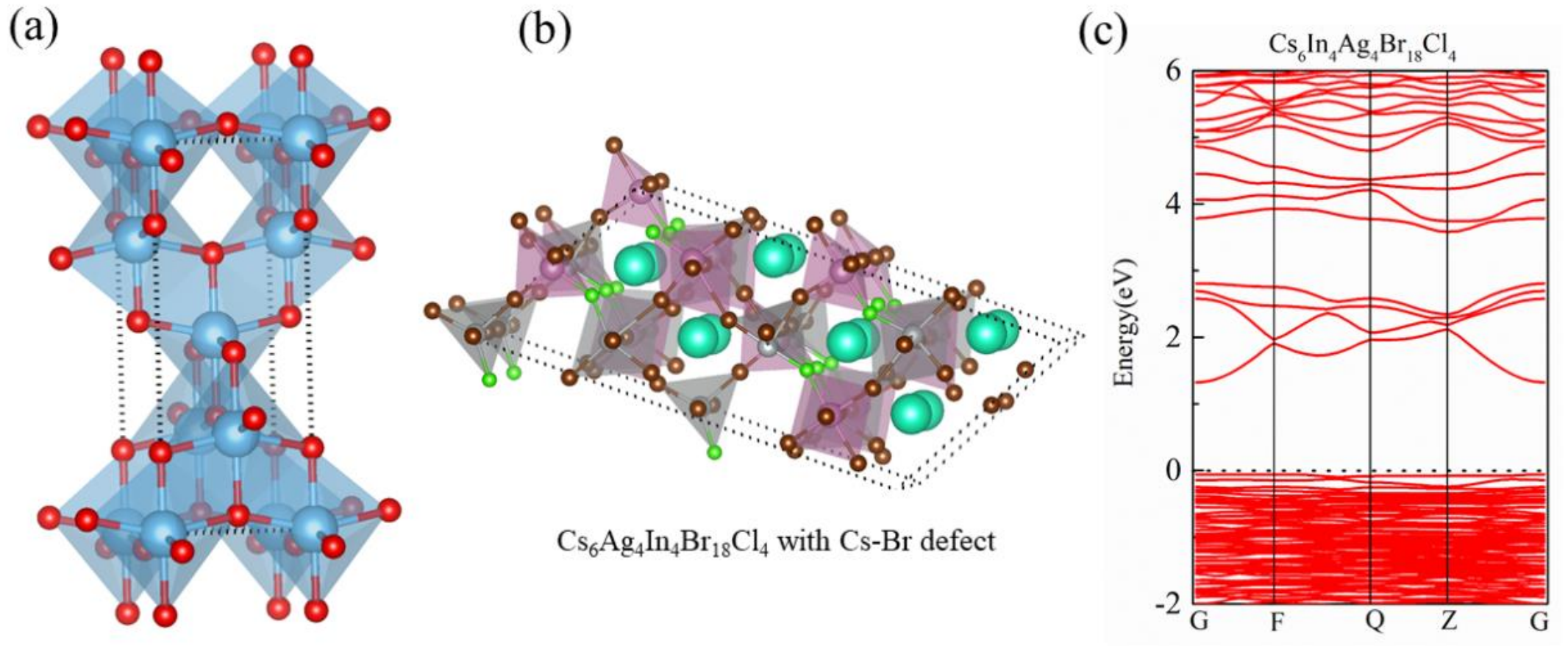

Fig.S2. (a) Crystal structures of anatase; (a) crystal structures of $Cs_6Ag_4In_4Br_{18}Cl_4$ with Cs-Br defect and PBE band structure of $Cs_6Ag_4In_4Br_{18}Cl_4$.